\documentclass{article}

\usepackage[english]{babel}
\usepackage{verbatim}
\usepackage[letterpaper,top=2cm,bottom=2cm,left=3cm,right=3cm,marginparwidth=1.75cm]{geometry}
\usepackage{amsmath,amsthm,amsfonts}
\usepackage{graphicx}
\usepackage[colorlinks=true, allcolors=blue]{hyperref}
\usepackage{bm}
\usepackage{parskip}
\usepackage{algorithm,algpseudocode}
\usepackage{tikz}
\usepackage{pgfplots}
  \pgfplotsset{compat = newest}
\usepackage[shortlabels]{enumitem}
\usepackage{xcolor}
\usepackage{qtree}
\usepackage{comment}
\usepackage{soul}
\usepackage[colorinlistoftodos]{todonotes}
\usepackage{nameref}
\usepackage{dirtree}
\usepackage{comment}
\usepackage{fancyvrb}
\usepackage{cite}
\usepackage{caption}
\usepackage{prooftrees}%
\usepackage[justification=centering]{caption}%

\newtheoremstyle{break}
  {10pt}
  {10pt}
  {\itshape}
  {}
  {\bfseries}
  {.}
  {\newline}
  {}

\theoremstyle{break}
\newtheorem{thm}{Theorem}[section]
\usepackage{float}
\usepackage{subcaption}
\usepackage{svg}
\usepackage{cleveref}
\usepackage{forest}
\usepackage{adjustbox}

\newcommand{\RR}{\mathbb{R}}

\newcommand{\QQ}{\mathbb{Q}}
\newcommand{\F}{\mathcal{F}}

\DeclareMathOperator{\lc}{lc}
\DeclareMathOperator{\tc}{tc}

\DeclareMathOperator{\res}{res}
\DeclareMathOperator{\disc}{disc}

\DeclareMathOperator{\Proj}{proj}

\DeclareMathOperator{\CAD}{CAD}

\definecolor{Green}{rgb}{0, 0.5, 0}

\newtheorem{definition}{Definition}

    \newcommand{\innerthmname}{}%

\title{Cylindrical Algebraic Decomposition in \textit{Macaulay2}}
\date{} %
\author{Corin Lee, Tereso del R\'{i}o and Hamid Rahkooy}

\begin{document}
\maketitle

\begin{abstract} 
\texttt{CylindricalAlgebraicDecomposition.m2} is the first implementation of Cylindrical Algebraic Decomposition (CAD) in \textit{Macaulay2}. CAD decomposes space into `cells' where input polynomials are sign-invariant. This package computes an Open CAD (full-dimensional cells only) for sets of real polynomials with rational coefficients, enabling users to solve existential problems involving strict inequalities. With the construction of a full CAD (cells of all dimensions), this tool could be extended to solve any real quantifier elimination problem. The current implementation employs the Lazard projection and introduces a new heuristic for choosing the variable ordering. 
\end{abstract}

\renewcommand\thefootnote{}%
\footnotetext{ \vspace{-3mm} 
\texttt{CylindricalAlgebraicDecomposition} version 1.0.3
}
\renewcommand\thefootnote{\arabic{footnote}}%

\section{Introduction}

This article documents the \texttt{CylindricalAlgebraicDecomposition} package for \textit{Macaulay2} \cite{M2}. Of particular note is the function \texttt{findPositiveSolution}, which allows the user to solve satisfiability problems on non-linear real arithmetic through the use of Open Cylindrical Algebraic Decomposition (Open CAD).

Cylindrical Algebraic Decomposition (CAD) is an important algorithm in symbolic computation for studying real semi-algebraic sets. It takes a set of multivariate polynomials and decomposes the solution space into disjoint regions known as \textit{cells}, within which the initial polynomials do not change sign. In satisfiability problems, this reduces the search from the uncountable real space to a finite number of regions.

CAD was first introduced as a method for performing \textit{quantifier elimination} (QE) over the reals in \cite{Collins75} and has applications in algebraic geometry and fields such as robotics, economics, and biology. CAD is well suited for computation and has been implemented in many widely used computer algebra packages, such as \textit{REDUCE, Maple} and \textit{Mathematica}.

However, CAD is computationally expensive, with a worst-case complexity that is doubly exponential in the number of variables \cite{DavenportHeintz87}. This package reduces the complexity of the original CAD algorithm proposed by Collins, incorporating modern improvements. It contains an implementation of Open CAD (see \Cref{sec:Open CAD}), uses the contemporary Lazard projection \cite{Lazard94,McCallumEtAl19} (see \Cref{sec:lazardprojection}), and is the first implementation of the \texttt{gmods} heuristic for variable ordering \cite{delRioEngland22} (see \Cref{sec:variable ordering}).

\subsection{Motivating Example}

Imagine we are interested in determining whether there exists an $x\in\RR$ such that \begin{equation}\label{eq:example use 1DCAD}
    g(x)=3-x^2>0\text{ \quad and \quad } h(x)=(7x-12)(x^2+x+1)>0.
\end{equation}
Studying the problem numerically, one could sample 1001 equispaced points between $-50$ and $50$ and find no point that satisfies the conditions. However, such a point does exist:

\begin{figure}[H]
    \centering
    \begin{subfigure}{0.49\textwidth}
        \centering
            \includegraphics[width=\textwidth]{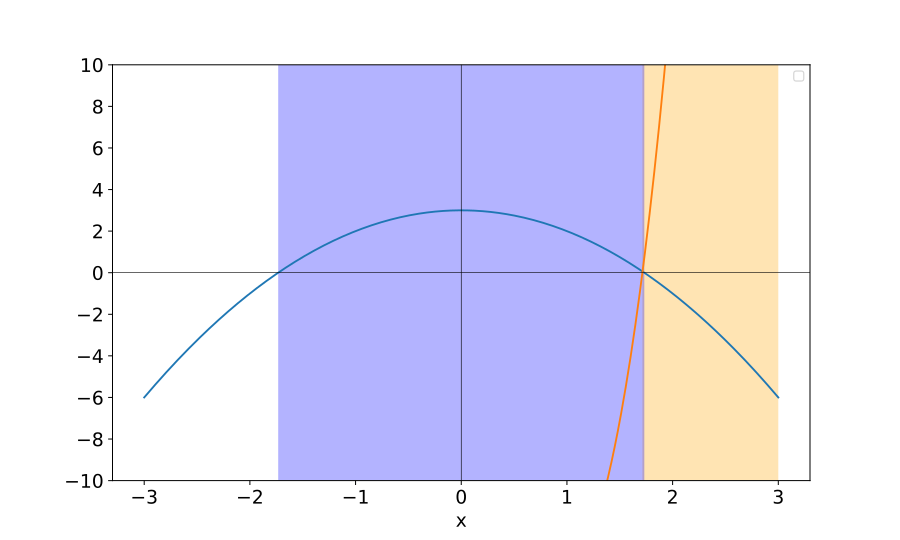}%
        \caption{It is not initially clear whether there is some value of $x$ that ensures both polynomials are positive...}
        \label{fig:closeplot-1}
    \end{subfigure}
    \begin{subfigure}{0.49\textwidth}
        \centering
            \includegraphics[width=\textwidth]{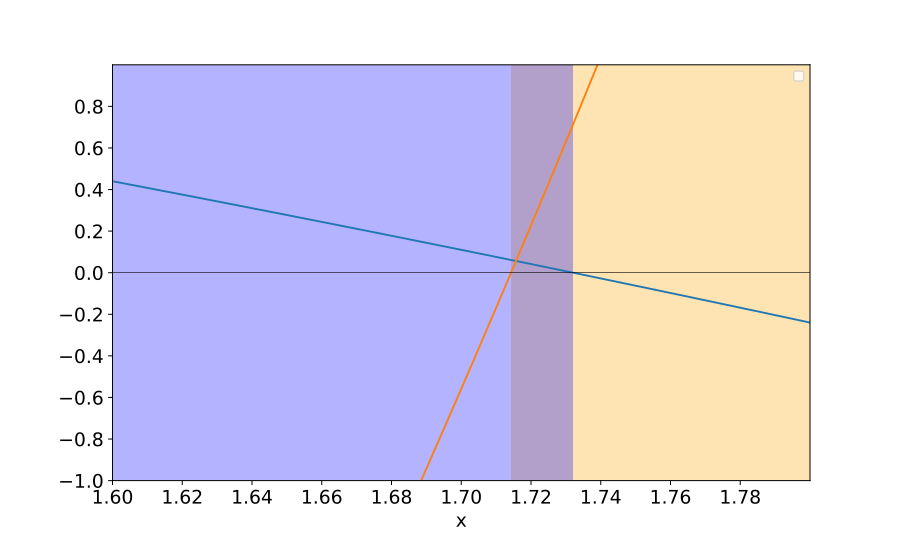}%
        \caption{...however it becomes clear that such an $x$ does exist when looking in greater detail.}
        \label{fig:closeplot-2}
    \end{subfigure}
    \caption{Algebraic varieties of $3-x^2$ (blue) and $(7x-12)(x^2+x+1)$ (orange), and the associated values of $x$ ensuring these polynomials are positive (blue and orange highlighting, respectively).}
    \label{fig:both-plots}
\end{figure}

The function \texttt{findPositiveSolution} in this package can check whether a set of strict polynomial inequalities has a solution. For the univariate example above, this function outputs the following:

\begin{Verbatim}[frame=leftline, framerule=4pt]
i1 : R=QQ[x]

i2 : findPositiveSolution({3-x^2,(7*x-12)*(x^2+x+1)})

                           1539
o2 = (true, HashTable{x => ----})
                            896
\end{Verbatim}

The package is not restricted to the univariate case and can find solutions for polynomials in as many variables as desired, given enough time.

\section{Cylindrical Algebraic Decomposition}\label{sec:CAD}

\subsection{Brief history}

A quantifier elimination (QE) problem involves turning a \textit{quantified formula} consisting of polynomial equations and inequalities combined with logical connectives such as $\land, \lor, \neg$, and quantifiers $\forall$ and $\exists$, into an equivalent \textit{quantifier-free formula}.

For example, the quantified formula $$(\exists x) \ (ax^2+bx+c=0) \land (a \neq 0) \text{ with } a, b, c\in\RR$$ is logically equivalent to $$b^2-4ac \geq 0.$$

In the 1940s, Tarksi showed through a constructive proof that real quantifier elimination is always possible \cite{Tarski51}. However, the algorithm used had complexity that could not be bounded by a finite tower of exponents, and so was considered impractical.

In 1975, Collins \cite{Collins75} proposed CAD as an algorithm that could solve real quantifier elimination problems with doubly exponential complexity in the number of variables \cite{BrownDavenport2007}. This complexity meant that the algorithm could only solve simple problems, but advances in computational power and CAD theory have allowed us to solve more challenging cases.

\subsection{CAD basics}

This paper covers the key parts of CAD, for a more detailed explanation of CAD see \cite{Jirstrand95}.

\begin{itemize}
    \item A \textbf{cell} is a connected region of $\RR^n$.
    
    \item A \textbf{decomposition} of a space $X$ is a collection of disjoint cells whose union is $X$.
    
    \item A cell is \textbf{cylindrical} with respect to the variable ordering $x_1\prec \dots \prec x_n$ if, for any $i<n$, it can be described as $$\{(x_1,\dots, x_n)\mid g_{i,1}(x_1,\dots, x_i)\leq x_{i+1} \leq g_{i,2}(x_1,\dots, x_i)\},$$ where $g_{i,1}$ and $g_{i,2}$ are continuous functions.
    
    \item Furthermore, a cell is \textbf{algebraic} if $g_{i,1}$ and $g_{i,2}$ are polynomials, and is called a \textit{$j$-cell} if it is homeomorphic to $\mathbb{R}^j$, for $0 \leq j \leq n$.
    
    \item A decomposition is \textbf{cylindrical} with respect to the variable ordering $x_1\prec \dots\prec x_n$ if, for any $i\leq n$, the projections of any two cells into $\RR[x_1, \dots, x_i]$ are either disjoint or identical.

    \item A CAD is a cylindrical decomposition whose cells are cylindrical and algebraic.
\end{itemize}

A CAD will only be useful if its cells satisfy some property of interest. We focus on \textit{sign-invariance}, where each input polynomial maintains a constant sign (positive, negative, or zero) within a cell. This invariance means it is sufficient to test a representative \textit{sample point} for each cell to determine the signs of the input polynomials.

\begin{definition}[Cylindrical Algebraic Decomposition]
    For $\mathcal{F}_n$ a set of polynomials in variables $x_1,\dots,x_n$, an $\mathcal{F}_n$-invariant CAD of $\RR^n$ (or colloquially `a CAD of $\mathcal{F}_n$'), denoted $\CAD(\mathcal{F}_n)$, is a cylindrical decomposition of $\RR^n$  whose cells are cylindrical and algebraic, and over which the polynomials in $\mathcal{F}_n$ have constant sign.
\end{definition}

\begin{figure}[h]
\centering
\begin{subfigure}[c]{.35\textwidth}
  \centering
  \includegraphics[width=.85\linewidth]{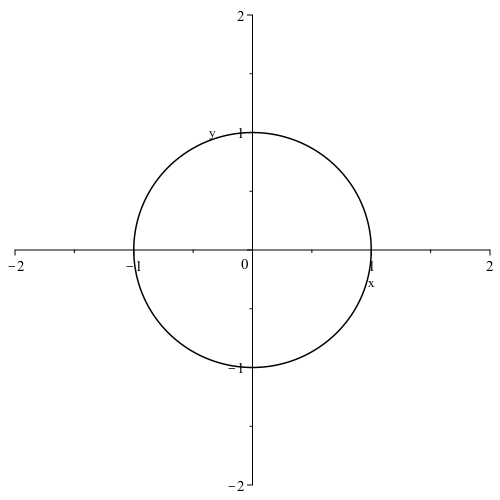}
\end{subfigure}%
\begin{subfigure}[c]{.05\textwidth}
  \centering
  \huge$\rightarrow$
\end{subfigure}%
\begin{subfigure}[c]{.35\textwidth}
  \centering
  \includegraphics[width=.85\linewidth]{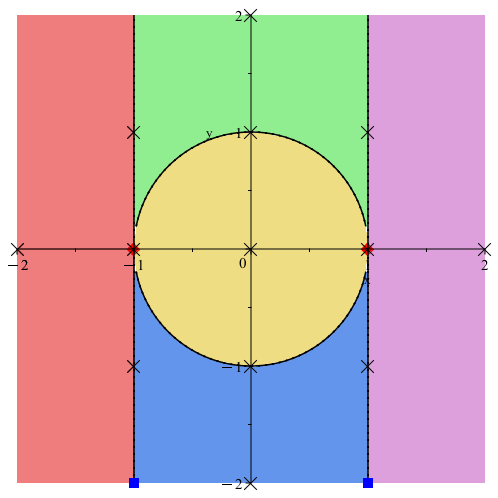}
\end{subfigure}
\caption{The graph of $\mathcal{F}=\{x^2+y^2-1\}$ and its associated sign-invariant CAD of $\RR^2$. This CAD consists of 13 cells: five 2-cells (the coloured regions), six 1-cells (the lines between them) and two 0-cells (the two red points where the lines meet). The black crosses represent the sample points for each cell. These cells stack in \textit{cylinders} over the five cells of the CAD of $\RR^1$ (the two points at the bottom and the regions between them). Each cell is described by constraints on $x$ and $y$.}
\label{fig:CadExample}
\end{figure}

For example, the collection of cells in Figure \ref{fig:CadExample} form a CAD of $\mathcal{F}=\{x^2+y^2-1\}$, with sample points $(-2,0), (-1,1), (-1,0), (-1,-1), (0,2), (0,1), (0,0), (0,-1), (0,-2), (1,1), (1,0), (1,-1)$ and $(2,0)$.

\subsection{CAD construction}

\noindent We will also use the term `CAD' to refer to the algorithm that generates a sign-invariant CAD for a given set of polynomials. Collins' original algorithm (sometimes referred to as a \textit{projection-and-lifting} CAD algorithm) uses a \textit{projection operator} on the input polynomials to produce a set of polynomials in one fewer variable, whose properties enable a CAD of the input polynomials to be constructed from a CAD of their projection.

In one dimension, constructing a sign-invariant CAD is straightforward: its cells are the collection of roots of the polynomials and the regions either side of them. Each sign-invariant cell defines a region where the number of real roots (the `root structure') does not change. 

In higher dimensions, we treat the polynomials in $\mathcal{F}_n$ as univariate in the greatest variable with respect to the ordering, written as $\mathcal{F}_n \subset \RR[x_1,\dots,x_{n-1}][x_n]$. To find changes in the root structure in $x_n$, we construct a set of \textit{projection polynomials} $\mathcal{F}_{n-1}$ in one fewer variable, whose roots indicate when changes occur (referred to as `projecting with respect to $x_n$', or `projecting away' $x_n$). This process is repeated, with each $\F_k$ obtained by projecting $\F^{k+1}$ with respect to $x_{k+1}$, until we reach $\mathcal{F}_{1} \subset \RR[x_1]$ whose roots and intervals make up a CAD of $\RR^1$. 

Due to the nature of the projection polynomials, the root structure of $\mathcal{F}_k$ does not vary over a cell of the CAD of $\RR^{k-1}$. We can continue the process and decompose the space above the cell into a new set of sign-invariant cells. The full collection of these cells forms the CAD of $\RR^k$.

\begin{center}
\begin{minipage}{0.5\textwidth}
    \noindent This algorithm follows a two-phase approach: 
    \begin{itemize}
        \item \textbf{Projection:}
        Repeatedly apply the projection operator to produce a chain of projection polynomial sets, each in one variable fewer than the previous, until reaching a univariate set of polynomials.
        \item \textbf{Lifting:} Starting with the univariate set, construct a CAD of the polynomials. Above this, one can create a new CAD in one dimension greater via \textit{lifting}, made possible by Theorem \ref{thm:collins_main_theorem}.
    \end{itemize}
    This recursive approach ultimately leads to a complete CAD for the original polynomial set.
\end{minipage}
\begin{minipage}{0.45\textwidth}
    \renewcommand{\arraystretch}{1.2}
    \[
    \begin{array}{ccc}
       \mathcal{F}_n \phantom{\! \Proj}     &  & \CAD(\mathcal{F}_{n}) \phantom{ \text{\scriptsize{Lift}}}\\
       \big\downarrow \! \scriptstyle\Proj  &  & \big\uparrow  \text{\scriptsize{Lift}} \\
       \mathcal{F}_{n-1} \phantom{\! \Proj} &  & \CAD(\mathcal{F}_{n-1}) \phantom{ \text{\scriptsize{Lift}}}\\
       \big\downarrow \! \scriptstyle\Proj  &  & \big\uparrow  \text{\scriptsize{Lift}} \\
       \ \vdots \phantom{\! \Proj}          &  & \vdots \phantom{ \text{\scriptsize{Lift}}}\\
       \big\downarrow \! \scriptstyle\Proj  &  & \big\uparrow  \text{\scriptsize{Lift}} \\
       \mathcal{F}_2 \phantom{\! \Proj}     &  & \CAD(\mathcal{F}_2) \phantom{ \text{\scriptsize{Lift}}}\\
       \big\downarrow \! \scriptstyle\Proj  &  & \big\uparrow  \text{\scriptsize{Lift}} \\
       \mathcal{F}_1 \phantom{\! \Proj}     & \hspace{-2.75em} \xrightarrow[\text{split into cells}]{\text{find roots}} \hspace{-1em} &  \CAD(\mathcal{F}_1) \phantom{ \text{\scriptsize{Lift}}}
    \end{array} 
    \]
\end{minipage}
\end{center}

\begin{thm}[\!\! \cite{Collins75}, Theorem 5]\label{thm:collins_main_theorem}
  Let $\mathcal{F}$ be a non-empty set of non-zero real polynomials in $n\geq 2$ real variables. Let $S$ be a connected subset of $\RR^{n-1}$. If every element of the projection of $\F$ is sign-invariant on $S$, then the polynomials in $\mathcal{F}$ are delineable over $S$.
\end{thm}

\begin{figure}[h]
    \begin{center}
    \includegraphics[width=0.7\linewidth]{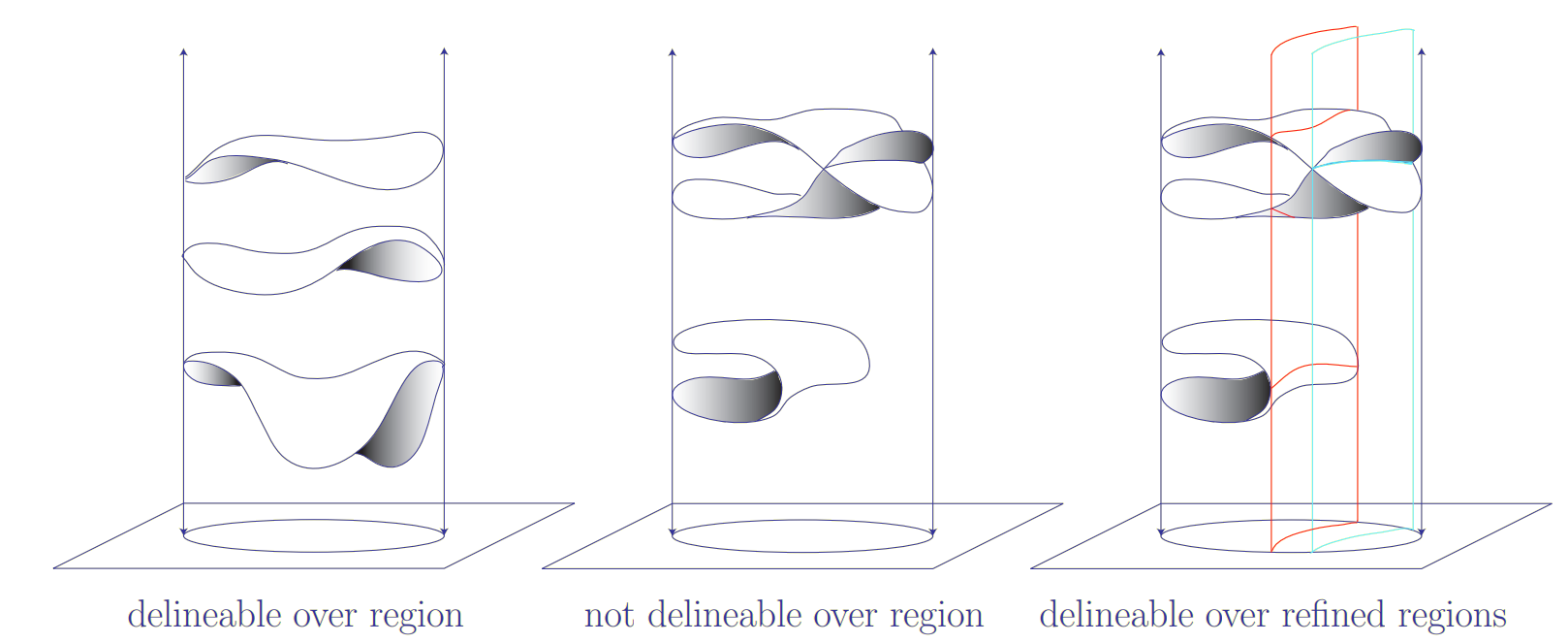}
    \caption{Illustration of delineability, from \cite{Brown2004}.}
    \label{fig:delineable}
    \end{center}
\end{figure}

\begin{definition}\label{def:delineability}
    A set of polynomials is \textbf{delineable} over $S$ if the number of distinct collective roots remains constant, i.e., their roots do not intersect or disappear (see Figure \ref{fig:delineable}).
\end{definition}

This theorem allows us to obtain an $n$-dimensional sign-invariant CAD by studying a finite set of sample points for the cells in an $(n-1)$-dimensional decomposition. For example, the sign-invariant CAD of $\mathcal{F}=\{x^2+y^2-1\}$ seen in Figure \ref{fig:CadExample} can be obtained by studying a point over each sign-invariant cell of the CAD of the projection polynomials $\{x-1,x+1\}$.

\subsection{Choices made for the package}

Certain decisions have been made in designing the CAD algorithm to best suit this package. These decisions are explained below.

\subsubsection{Variable ordering}\label{sec:variable ordering}

\begin{figure}[h]
	\centering
	\includegraphics[width=0.48\textwidth]{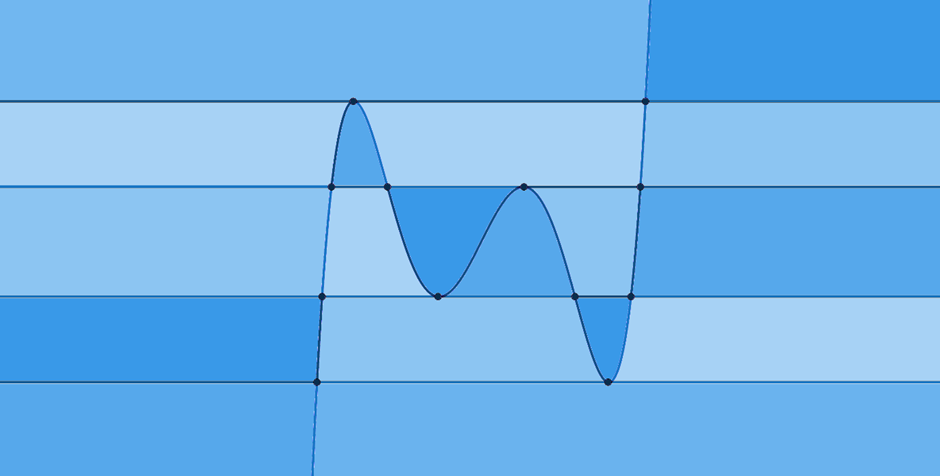}
    \hspace{0.1pt}
	\includegraphics[width=0.48\textwidth]{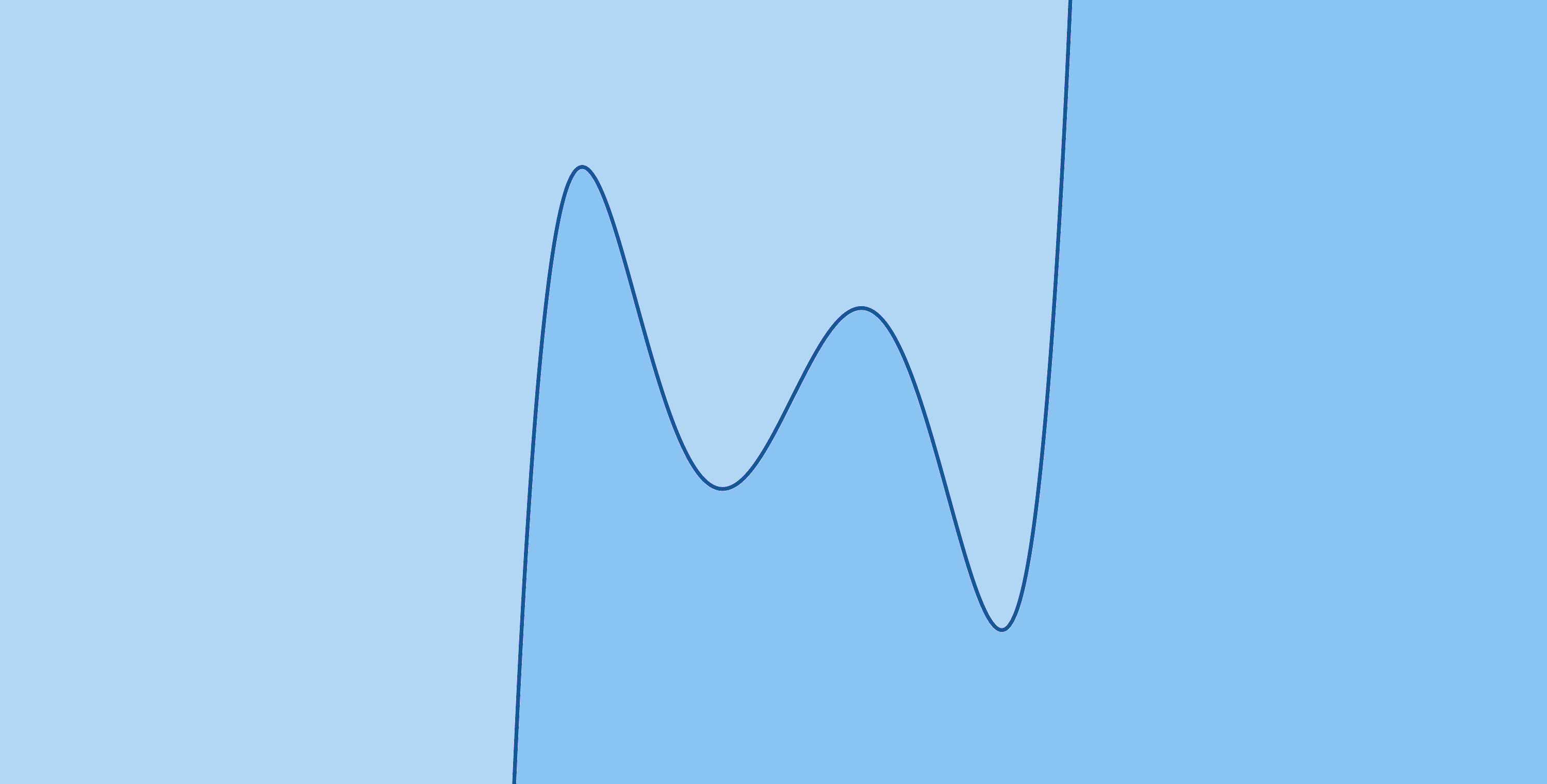}
	\caption{CADs sign-invariant for the set of polynomials $\{x^5+5 x^4+5 x^3-5 x^2-6x-2y\}.$
            Using ordering $y \prec x$, we obtain a CAD with 57 cells (18 2-cells, 27 1-cells and 12 0-cells).
            Using the ordering $x \prec y$ generates only three cells (two 2-cells and one 1-cell).  
            \label{fig: importanceordering}
            }
\end{figure}

In order to compute a CAD, one must decide on the variable ordering. The ordering chosen can significantly impact the number of cells in a CAD (see Figure \ref{fig: importanceordering}), and thus the resources and time needed to produce it. Considerable work has been done trying to find the best way to choose the variable ordering \cite{HuangEtAl2015, EnglandFlorescu2019, ChenEtAl2020}.

\texttt{CylindricalAlgebraicDecomposition} employs \texttt{gmods}, a strategy developed by Del Río and England \cite{delRioEngland22}, which first projects the variable with the lowest degree sum across the set of polynomials.

\subsubsection{Lazard projection}\label{sec:lazardprojection}

This package uses the \textit{Lazard projection}, originally proposed by Lazard \cite{Lazard94}, later validated by McCallum et al. \cite{McCallumEtAl19}. Constructing an open CAD eliminates the need for extra checks that would otherwise be needed when applying the Lazard projection in a full CAD. %

\begin{definition}[Lazard projection]\label{def:lazardprojection}

When projecting, we consider polynomials in $\RR[x_1,\dots,x_n]$ as univariate in $x_n$ with coefficients in $\RR[x_1,\dots,x_{n-1}]$, hence the notion of `leading coefficient' is in terms of $x_n$.

Let $\mathcal{F}_n$ be a finite set of irreducible pairwise coprime polynomials in $\RR[x_1,\dots,x_n]$, with $n \geq 2$. The \textit{Lazard projection} of $\mathcal{F}_n$, $P_L(\mathcal{F}_n) \subset \RR[x_1,x_2,\dots,x_{n-1}]$, comprises the following polynomials:

\begin{enumerate}
    \item all leading coefficients of the elements of $\mathcal{F}_n$,
    \item all trailing coefficients (coefficients independent of $x_n$) of the elements of $\mathcal{F}_n$,
    \item all discriminants of the elements of $\mathcal{F}_n$, and
    \item all resultants of pairs of distinct elements of $\mathcal{F}_n$.
\end{enumerate}

$$P_L(\mathcal{F}_n) = \{\lc(f_i)\} \cup \{\tc(f_i)\} \cup \{\disc(f_i)\} \cup \{\res(f_i,f_j \mid i \neq j)\}.$$

\end{definition}

The roots of these projection polynomials sufficiently capture the points at which the root structure of $\mathcal{F}_n$ changes, such as when two distinct roots meet or the polynomial degree drops. In \texttt{CylindricalAlgebraicDecomposition}, the set of input and output polynomials are replaced by  their nonconstant irreducible factors, ensuring they are irreducible and pairwise coprime (see \ref{sec:lazardprojection}).

\subsubsection{Open CAD}\label{sec:Open CAD}

We focus on the notion of an \textit{Open CAD}, that is, open solution sets of systems of strict polynomial inequalities \cite{Strzebonski00}. In other words, an Open CAD will only consist of open, full-dimensional $n$-cells.

At present, \textit{Macaulay2} does not support polynomial division over the reals or symbolic root storage, making a full CAD impossible due to the inability to store or lift over algebraic irrational roots. Despite this limitation, many real-world applications are not interested in these zero-measure regions, so an Open CAD is enough for most cases.

As a result, we have implemented an Open CAD algorithm, using the existing \texttt{RealRoots} package \cite{LopezGarcia2024} to isolate roots within rational intervals, allowing rational sample points to be chosen which represent every open cell. Additionally, inefficiencies and issues within the \texttt{RealRoots} package have been corrected and improved.%

\subsection{\texttt{RealRoots} improvements}

\texttt{CylindricalAlgebraicDecomposition} calls the existing \texttt{RealRoots} package to perform real root isolation. However, there are some limitations and inefficiencies in \texttt{RealRoots} that we have addressed:
\small
\begin{itemize}
    \item \textbf{Only works on one polynomial:}  \texttt{CylindricalAlgebraicDecomposition} simply performs this real root isolation on $\prod{f_i}$, where each $f_i$ is reduced to its irreducible factors during the projection phase.
    \item \textbf{Error in root bound:} The original root bound failed when the leading coefficient was negative. This was later fixed in the JSAG release of the package, but has also been fixed here.
    \item \textbf{Inefficient root bound:} The original root bound was weaker than necessary, and would scale badly in cases of coefficient blowup. The fix improves the existing bound, implements another which better handles large coefficients, and the command now takes the smaller bound, thus speeding up each real root isolation step.
    \item \textbf{Inefficient with close roots:} The original algorithm would continue to refine every interval in cases when one root required a smaller interval, creating extra work. Now, once a root has been sufficiently isolated, its interval is left alone.
    \item \textbf{Roots not always ordered:} Root intervals are now consistently listed lowest to highest.
\end{itemize}

\section{The \texttt{CylindricalAlgebraicDecomposition} package}

This section provides a detailed description of the \texttt{CylindricalAlgebraicDecomposition}. This package allows the user to produce an open, sign-invariant CAD with respect to a list of input polynomials in $\QQ[x_1,\dots,x_n]$ by constructing a tree of sample points in $\QQ^n$, representing the open cells of a CAD of $\RR^n$. It also allows users to step through each phase of the Open CAD algorithm and check for points where all input polynomials are positive.

\subsection{Package Overview}

\begin{figure}[htbp]
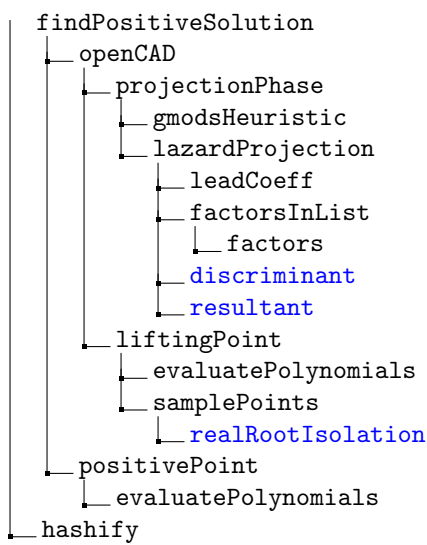

    \centering
    \begin{minipage}{.1\textwidth}
        \hspace{0cm}
    \end{minipage}%
    \begin{minipage}{.3\textwidth}
        \dirtree{%
.1 findPositiveSolution.
.2 openCAD.
.3 projectionPhase.
.4 gmodsHeuristic.
.4 lazardProjection.
.5 leadCoeff.
.5 factorsInList.
.6 factors.
.5 \textcolor{blue}{discriminant}.
.5 \textcolor{blue}{resultant}.
.3 liftingPoint.
.4 evaluatePolynomials.
.4 samplePoints.
.5 \textcolor{blue}{realRootIsolation}.
.2 positivePoint.
.3 evaluatePolynomials.
.1 hashify.
}
    \end{minipage}
    \captionsetup{justification=centering}
    \caption{List of functions contained in the \texttt{CylindricalAlgebraicDecomposition} package. \\ Methods labelled in blue are from external packages.}
\end{figure}

The package's top-level function is \texttt{findPositiveSolution}, which takes a list of input polynomials and returns a point in which all input polynomials are strictly positive or informs the user that no such point exists.

This is built upon two other functions: \texttt{openCAD}, which constructs an Open CAD for the given list of polynomials as a tree of sample points, and \texttt{positivePoint} which takes the tree (or a branch with leaves) and checks if such a positive point exists. These commands are also available to the user, allowing them to construct Open CADs and check the cells as needed.

Now we proceed to describe these functions in more detail before continuing with the remaining functions in the package, providing inputs, outputs, and examples. To illustrate the algorithm, we will use the following system of polynomials $\mathcal{F}=\{f_1,f_2\}$, taken from \cite{Jirstrand95}:

\begin{equation}\label{eq:JirstrandExample}
\begin{array}{ll}
   f_1 =& x_1^2+x_2^2-1\\
   f_2 =& x_1^3-x_2^2.
\end{array} 
\end{equation}

\begin{Verbatim}[frame=leftline, framerule=4pt]
R=QQ[x1,x2];
f1:=x1^2+x2^2-1; f2:=x1^3-x2^2;
F={f1,f2};
\end{Verbatim}

\subsubsection{\texttt{openCAD}} \label{sec:opencad_command}

\textbf{Input:} List of polynomials $\mathcal{F}_n=\{f_{1},\dots,f_{r}\}$ in variables $x_1,\dots,x_n$.

\textbf{Output:} Tree of sample points representing the Open CAD of $\RR^n$ with respect to $\mathcal{F}_n$.

The command \texttt{openCAD} produces an open, projection-and-lifting CAD, and consists of a projection phase (see \Cref{sec:projectionPhase}) and a lifting phase (see \Cref{sec:liftingPhase}).

The format of \texttt{openCAD} is a tree, represented in \textit{Macaulay2} as a collection of nested hash tables with $n$ levels. Each level $k$, for $1 \leq k < n$, represents the CAD of $\RR^k$, containing the projection polynomials $\mathcal{F}_k$, the cell below (as a $(k-1)$-dimensional sample point) and the collection of sample points in $x_k$ obtained by lifting over that cell. Each leaf node (level $n$) represents an $n$-dimensional sample point of $\mathcal{F}_n$.

\begin{Verbatim}[frame=leftline, framerule=4pt]
i3 : openCAD(F)

o3 = MutableHashTable{...7...}
\end{Verbatim}

The output of \texttt{openCAD} is substantial, and thus is left unprinted in order to minimise clutter (see \ref{sec:hashify} for the full output). In this example, the first level consists of $\mathcal{F}_1$ (\texttt{polynomials}), the (empty) cell below (\texttt{point}), and the five open sample points of the CAD of $\RR^1$.

\begin{footnotesize}
\begin{Verbatim}[frame=leftline, framerule=4pt]
i4 : peek(openCAD(F))

o4 = MutableHashTable{point => MutableHashTable{}                       }
                                                            3     2
                      polynomials => {x1 - 1, x1 + 1, x1, x1  + x1  - 1}
                        3
                      - - => MutableHashTable{...5...}
                        4
                        5
                      - - => MutableHashTable{...2...}
                        2
                      3
                      - => MutableHashTable{...7...}
                      8
                      15
                      -- => MutableHashTable{...7...}
                      16
                      17
                      -- => MutableHashTable{...5...}
                       8
\end{Verbatim}
\end{footnotesize}

\subsubsection{\texttt{positivePoint}}

\textbf{Input 1:} List of polynomials $\mathcal{F}_n=\{f_{1},\dots,f_{r}\}$ in variables $x_1,\dots,x_n$. \\
\textbf{Input 2:} A tree of sample points representing an Open CAD (or a subtree reaching down to level $n$).

\textbf{Output:} If every input polynomial is positive at an $n$-dimensional sample point, that sample point is returned. If not, `\textit{no point exists}' is returned.

\begin{Verbatim}[frame=leftline, framerule=4pt]
i5 : positivePoint(F,openCAD(F))

o5 = MutableHashTable{...2...}
\end{Verbatim}

Yes, such a point exists. The point is:

\begin{Verbatim}[frame=leftline, framerule=4pt]
i6 : peek positivePoint(F,openCAD(F))

o6 = MutableHashTable{x2 => 0 }
                            17
                      x1 => --
                             8
\end{Verbatim}

\subsubsection{\texttt{findPositiveSolution}}

\textbf{Input:} List of polynomials $\mathcal{F}_n=\{f_{1},\dots,f_{r}\}$ in variables $x_1,\dots,x_n$.

\textbf{Output:} Returns a boolean indicating if there exists a point in which all polynomials are strictly positive, and if so, it returns one such point as a hash table.

The command \texttt{findPositiveSolution} combines the previous commands, first producing an Open CAD from the input polynomials, then checking this for a point where the input polynomials are positive.

\begin{Verbatim}[frame=leftline, framerule=4pt]
i7 : findPositiveSolution(F)

                            17
o7 = (true, HashTable{x1 => --})
                             8
                      x2 => 0                             
\end{Verbatim}

\begin{figure}[H]
    \centering
    \includegraphics[width=0.5\linewidth]{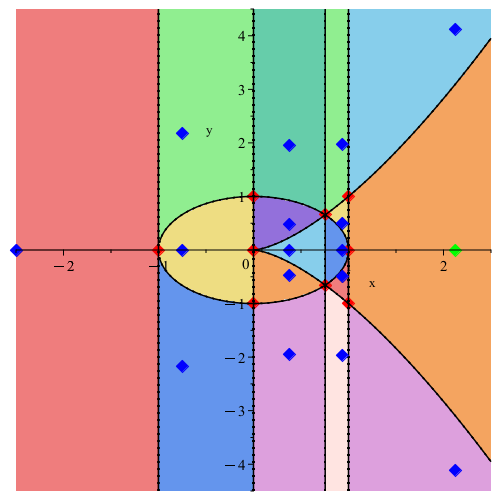}

\begin{center}
    \scalebox{0.95}{
    \scriptsize
    \begin{forest}                      
    for tree={
        grow=south,
        anchor=north,
        draw=none,
        edge={->},
        l sep=5em,
        s sep=0.2em,
    },
    [,phantom                           %
        [,phantom,for tree={no edge}
            [\boldsymbol{$x_1: \qquad$}, tier=b                            
                [\boldsymbol{$x_2: \qquad$},  tier=c]
            ]
        ]
        [$\bullet$, no edge
            [$-\frac{5}{2}$,tier=b
            	[$0$, tier=c]
            ]
            [$-\frac{3}{4}$,tier=b
            	[$-\frac{2\,237}{1\,024}$, tier=c]
            	[$0$, tier=c]
            	[$\frac{2\,237}{1\,024}$, tier=c]
            ]
            [$\frac{3}{8}$,tier=b
                [$-\frac{2\,003}{1\,024}$, tier=c]	
                [$-\frac{979}{2\,048}$, tier=c]
            	[$0$, tier=c]
            	[$\frac{979}{2\,048}$, tier=c]
            	[$\frac{2\,003}{1\,024}$, tier=c]
            ]
            [$\frac{15}{16}$,tier=b
            	[$-\frac{16\,159}{8\,192}$, tier=c]
            	[$-\frac{7\,967}{16\,384}$, tier=c]
            	[$0$, tier=c]
            	[$\frac{7\,967}{16\,384}$, tier=c]
            	[$\frac{16\,159}{8\,192}$, tier=c]
            ]
            [\textcolor{Green}{$\frac{17}{8}$},tier=b, edge=Green
            	[$-\frac{33}{8}$, tier=c]
            	[\textcolor{Green}{$0$}, tier=c, edge=Green, draw=green]
            	[$\frac{33}{8}$, tier=c]
            ]
        ]
    ]
    \end{forest}
   }
\end{center}
\caption{\textbf{Top:} A visualisation of a (full) CAD of $\mathcal{F}$. The coloured regions represent the open cells, the green diamond represents the sample point from \texttt{positivePoint}/\texttt{findPositiveSolution}, and the blue diamonds represent the remaining sample points from \texttt{openCAD}. \\ \textbf{Bottom:} A representation of these sample points in a tree structure.} \label{fig:graphandtree}
\end{figure}

\subsection{\texttt{hashify}}\label{sec:hashify}

\textbf{Input:} A data structure.

\textbf{Output:} The same data structure with any mutable hash tables converted to immutable ones.

The recursive nature of this Open CAD algorithm greatly benefits from the ability to copy and modify previous steps, something only possible with a \textit{mutable} hash table. Additionally, mutable hash tables do not display their entries by default, avoiding the otherwise overwhelming output given by \texttt{openCAD} and allowing the user to expand the parts of interest as they see fit.

However, mutable hash tables are incomparable and cannot be fully expanded. The \texttt{hashify} command was implemented to address this, converting mutable hash tables into immutable ones which are able to be viewed in their entirety.

Fully expanding the output of \texttt{openCAD} illustrates the structure of the computed Open CAD (compare this with Figure \ref{fig:graphandtree}), while also demonstrating that \texttt{hashify} should be used with care:

\begin{figure}[H]
\begin{Verbatim}[frame=leftline, framerule=4pt]
i8 : openCAD(F)

o8 = MutableHashTable{...7...}

i9 : hashify openCAD(F)                           
\end{Verbatim}
\ \\
\begin{adjustbox}{max width=0.5\linewidth}
\begin{BVerbatim}[frame=leftline, framerule=4pt]
                 3                2237                                         3
o9 = HashTable{- - => HashTable{- ---- => HashTable{point => HashTable{x1 => - -   }}} }
                 4                1024                                         4
                                                                               2237
                                                                       x2 => - ----
                                                                               1024
                                                                          3
                                0 => HashTable{point => HashTable{x1 => - -}}
                                                                          4
                                                                  x2 => 0
                                2237                                         3
                                ---- => HashTable{point => HashTable{x1 => - - }}
                                1024                                         4
                                                                           2237
                                                                     x2 => ----
                                                                           1024
                                                           3
                                point => HashTable{x1 => - -}
                                                           4
                                                  2    7      2   27
                                polynomials => {x2  - --, - x2  - --}
                                                      16          64
                 5                                                        5
               - - => HashTable{0 => HashTable{point => HashTable{x1 => - -}}}
                 2                                                        2
                                                                  x2 => 0
                                                           5
                                point => HashTable{x1 => - -}
                                                           2
                                                  2   21      2   125
                                polynomials => {x2  + --, - x2  - ---}
                                                       4           8
               3                 979                                       3
               - => HashTable{- ---- => HashTable{point => HashTable{x1 => -     }}}
               8                2048                                       8
                                                                              979
                                                                     x2 => - ----
                                                                             2048
                                2003                                       3
                              - ---- => HashTable{point => HashTable{x1 => -     }}
                                1024                                       8
                                                                             2003
                                                                     x2 => - ----
                                                                             1024
                                                                      3
                              0 => HashTable{point => HashTable{x1 => -}}
                                                                      8
                                                                x2 => 0
                               979                                       3
                              ---- => HashTable{point => HashTable{x1 => -   }}
                              2048                                       8
                                                                          979
                                                                   x2 => ----
                                                                         2048
                              2003                                       3
                              ---- => HashTable{point => HashTable{x1 => -   }}
                              1024                                       8
                                                                         2003
                                                                   x2 => ----
                                                                         1024
                                                       3
                              point => HashTable{x1 => -}
                                                       8
                                                2   55      2    27
                              polynomials => {x2  - --, - x2  + ---}
                                                    64          512
\end{BVerbatim}
\end{adjustbox}
\begin{adjustbox}{max width=0.5\linewidth}
\begin{BVerbatim}[frame=leftline, framerule=4pt]
               15                 7967                                       15
               -- => HashTable{- ----- => HashTable{point => HashTable{x1 => --     }}}
               16                16384                                       16
                                                                                7967
                                                                       x2 => - -----
                                                                               16384
                                 16159                                       15
                               - ----- => HashTable{point => HashTable{x1 => --     }}
                                  8192                                       16
                                                                               16159
                                                                       x2 => - -----
                                                                                8192
                                                                       15
                               0 => HashTable{point => HashTable{x1 => --}}
                                                                       16
                                                                 x2 => 0
                                7967                                       15
                               ----- => HashTable{point => HashTable{x1 => --   }}
                               16384                                       16
                                                                            7967
                                                                     x2 => -----
                                                                           16384
                               16159                                       15
                               ----- => HashTable{point => HashTable{x1 => --   }}
                                8192                                       16
                                                                           16159
                                                                     x2 => -----
                                                                            8192
                                                        15
                               point => HashTable{x1 => --}
                                                        16
                                                 2    31      2   3375
                               polynomials => {x2  - ---, - x2  + ----}
                                                     256          4096
               17                33                                       17
               -- => HashTable{- -- => HashTable{point => HashTable{x1 => --  }}}
                8                 8                                        8
                                                                            33
                                                                    x2 => - --
                                                                             8
                                                                       17
                               0 => HashTable{point => HashTable{x1 => --}}
                                                                        8
                                                                 x2 => 0
                               33                                       17
                               -- => HashTable{point => HashTable{x1 => --}}
                                8                                        8
                                                                        33
                                                                  x2 => --
                                                                         8
                                                        17
                               point => HashTable{x1 => --}
                                                         8
                                                 2   225      2   4913
                               polynomials => {x2  + ---, - x2  + ----}
                                                      64           512
               point => HashTable{}
                                                     3     2
               polynomials => {x1 - 1, x1 + 1, x1, x1  + x1  - 1}





               
\end{BVerbatim}
\end{adjustbox}
\end{figure}

We are now able to see the full structure of \texttt{openCAD}: each level of this tree corresponds to a variable, listing sample points based on those of the level below. In this example we can see the full list of open sample points of the CAD of $\F$. For a deeper explanation of this structure see \ref{sec:liftingPhase}

\normalsize
\color{black}

\subsection{Projection phase}\label{sec:projectionPhase}

The entire projection process is handled by the command \texttt{projectionPhase}, which takes the list of input polynomials $\F_n$ and returns the full collection of projection polynomial lists $\{\F_1,\dots,\F_n\}$, along with the corresponding variable ordering $x_1 \prec \dots \prec x_n$. At each step, \texttt{gmodsHeuristic} selects a variable to eliminate, and \texttt{lazardProjection} constructs the resulting projection polynomials.

\subsubsection{\texttt{gmodsHeuristic}}\label{sec:gmodsHeuristic}

\textbf{Input 1:} A list of polynomials $\mathcal{F}_n=\{f_{1},\dots,f_{r}\}$ in $n$ variables. \\
\textbf{Input 2:} The list of variables $x_1,\dots,x_n$ appearing in $\F_n$.

\textbf{Output:} The variable to project as suggested by the \texttt{gmods} heuristic (see \Cref{sec:variable ordering}).

\begin{Verbatim}[frame=leftline, framerule=4pt]
i10 : gmodsHeuristic(F,support(F))

o10 = x2
\end{Verbatim}

\subsubsection{\texttt{lazardProjection}} \label{sec:lazardProjection}

\textbf{Input 1:} List of polynomials $\mathcal{F}_n=\{f_{1},\dots,f_{r}\}$ in variables $x_1,\dots,x_n$. \\
\textbf{Input 2:} The variable to be projected away, typically the highest variable $x_n$.

\textbf{Output:} A list of projection polynomials $\mathcal{F}_{n-1}$ in $x_1,\dots,x_{n-1}$, obtained via the \textit{Lazard projection} (Definition \ref{def:lazardprojection}) with respect to $x_{n}$.

This command ensures polynomials are irreducible and pairwise coprime using the subcommand \texttt{factorsInList} to break them into their unique irreducible factors, discarding constants. The Lazard projection is then applied with respect to $x_n$, returning the list of factors of the leading and trailing coefficients, discriminants and resultants.

Projecting $\F$ with respect to $x_2$ returns:

\begin{Verbatim}[frame=leftline, framerule=4pt]
i11 : lazardProjection(F,x2)

                             3     2
o11 = {x1 - 1, x1 + 1, x1, x1  + x1  - 1}
\end{Verbatim}
where
\begin{itemize}
    \item $x_1 - 1$ and $x_1 + 1$ are factors of the trailing coefficient and discriminant of $f_1$,
    \item $x_1$ is a factor of the trailing coefficient of $f_2$, and
    \item $x_1^3  + x_1^2  - 1$ is the resultant of $f_1$ and $f_2$.
\end{itemize}

\subsubsection{\texttt{projectionPhase}}

\textbf{Input:} List of polynomials $\mathcal{F}_n=\{f_{1},\dots,f_{r}\}$ in variables $x_1,\dots,x_n$.

\textbf{Output 1:} List of projection polynomial lists $\{\F_1,\dots, \F_n\}$ (see \Cref{sec:CAD} for more details). \\
\textbf{Output 2:} The associated variable ordering $x_1 \prec \dots \prec x_n$.

\begin{Verbatim}[frame=leftline, framerule=4pt]
i12 : projectionPhase(F)

                               3     2          2     2        3     2
o12 = ({{x1 - 1, x1 + 1, x1, x1  + x1  - 1}, {x1  + x2  - 1, x1  - x2 }}, {x1,x2})
\end{Verbatim}

\subsection{Lifting phase}\label{sec:liftingPhase}

Like with the projection phase, the full lifting phase is handled by one command. Given a list of projection polynomial lists, their associated ordering, and a $(k-1)$-dimensional sample point representing the cell below, the command \texttt{liftingPoint} recursively constructs cells above each sample point until a full collection of $n$-dimensional sample points is obtained.

Ordinarily (as is the case for \texttt{openCAD}, see \ref{sec:opencad_command}), an empty (0-dimensional) sample point representing the root node of the tree is used, leading to a CAD of $\RR^n$. This command also allows the construction of a `sub-CAD' of the space above any provided sample point.

Each sample point is stored as a hash table mapping variables to values. The command \texttt{liftingPoint} calls \texttt{evaluatePolynomials} to evaluate polynomials above the sample point, returning a list of univariate polynomials. This is then passed to \texttt{samplePoints}, which isolates roots and returns the open sample points of the space.

\subsubsection{\texttt{samplePoints}}

\textbf{Input:} List of univariate polynomials $\mathcal{F}=\{f_{1},\dots,f_{r}\} \subset \RR[x]$.

\textbf{Output:} A list of open sample points for the CAD of the (one-dimensional) space containing $\F$.

This command applies \texttt{realRootIsolation} from \texttt{RealRoots} to the polynomials, isolating each real root within a rational interval, then returns sample points either side of these intervals. If applied to our projection polynomials $\F_1$, these are the open sample points of $\CAD(\F_1)$, but if applied to $\F_k$ evaluated at a $(k-1)$-dimensional sample point, these are the open sample points of the CAD of the space above that point.

For example, we can obtain the open sample points of our projection polynomials obtained in \ref{sec:lazardProjection}:

\begin{Verbatim}[frame=leftline, framerule=4pt]
i13 : samplePoints(lazardProjection(F,x2))

         5    3  3  15  17
o13 = {- -, - -, -, --, --}
         2    4  8  16   8
\end{Verbatim}

These are the values of $x_1$ seen in Figure \ref{fig:graphandtree}.

\subsubsection{\texttt{evaluatePolynomials}}

\textbf{Input 1:} List of polynomials $\mathcal{F}_n=\{f_{1},\dots,f_{r}\}$ in variables $x_1,\dots,x_n$. \\
\textbf{Input 2:} A $k$-dimensional point.

\textbf{Output:} The list of polynomials evaluated at this point.

Above one of these sample points, our new univariate polynomials become:

\begin{Verbatim}[frame=leftline, framerule=4pt]
alpha = new MutableHashTable; alpha#x1 = -3/4;

i14 : evaluatePolynomials(F,alpha)

         2    7      2   27
o14 = {x2  - --, - x2  - --}
             16          64
\end{Verbatim}

We can then look at the open sample points of $x_2$ above this:

\begin{Verbatim}[frame=leftline, framerule=4pt]
i15 : samplePoints(evaluatePolynomials(F,alpha))

         2237     2237
o15 = {- ----, 0, ----}
         1024     1024
\end{Verbatim}

\subsubsection{\texttt{liftingPoint}}

\textbf{Input 1:} List of projection polynomial lists $\{\F_1,\dots, \F_n\}$. \\
\textbf{Input 2:} The associated variable ordering $x_1 \prec \dots \prec x_n$. \\
\textbf{Input 3:} A $(k-1)$-dimensional point.

\textbf{Output:} An Open CAD (see \Cref{sec:CAD}) over the given point.

Above each sample point of the previous level, a cell is constructed. This level $k$ cell contains:

\begin{itemize}
    \item the $(k-1)$ dimensional sample point below (\texttt{point}),
    \item the polynomials of $\F_k$ evaluated at this sample point (\texttt{polynomials}),
    \item the values of $x_k$ for each $k$-dimensional sample point above \texttt{point}. The command \texttt{liftingPoint} is called again with these new sample points.
\end{itemize}

At the top level (level $n$), each cell only contains \texttt{point}, the sample point for that open cell.

Lifting above the previous sample point:
\begin{footnotesize}

\begin{Verbatim}[frame=leftline, framerule=4pt]
(PP,ord) = projectionPhase(F);

i16 : hashify(liftingPoint(PP,ord,alpha))

                  2237                                         3
o16 = HashTable{- ---- => HashTable{point => HashTable{x1 => - -   }}}
                  1024                                         4
                                                               2237
                                                       x2 => - ----
                                                               1024
                                                          3
                0 => HashTable{point => HashTable{x1 => - -}}
                                                          4
                                                  x2 => 0
                2237                                         3
                ---- => HashTable{point => HashTable{x1 => - - }}
                1024                                         4
                                                           2237
                                                     x2 => ----
                                                           1024
                                           3
                point => HashTable{x1 => - -}
                                           4
                                  2    7      2   27
                polynomials => {x2  - --, - x2  - --}
                                      16          64
\end{Verbatim}
\end{footnotesize}
This demonstrates the lifting phase: we first isolate their real roots of the univariate polynomials in $\F_1$ and select sample points from each open cell around the roots. These open sample points are then substituted into $\F_2$ to produce a new list of univariate polynomials, and the process repeats until a full Open CAD is constructed.

\section{Performance - Hyperspheres}

\begin{figure}[H]
    \centering
    \includegraphics[width=0.5\linewidth,trim={0 3cm 0 3cm},clip]{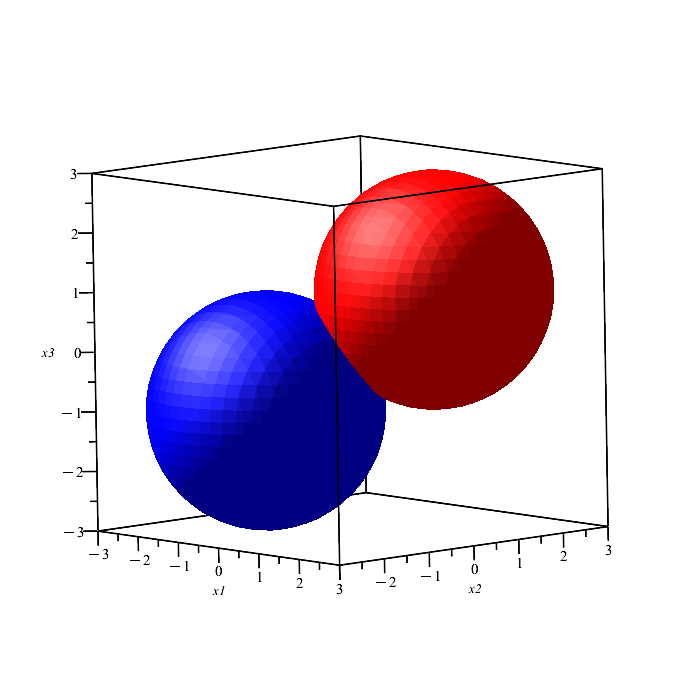}
    \caption{A plot of the algebraic variety of $\F_n$. \\Two intersecting spheres centered at (1,1,1) and (-1,-1,-1) respectively.}
    \label{fig:spheres}
\end{figure}

It is known that CAD has doubly exponential complexity, meaning that only problems in fewer variables can be solved. To measure the performance of our package, we created CADs for the following intersecting $n$-dimensional spheres (of which the three-dimensional case is depicted in \Cref{fig:spheres}):

$$\mathcal{F}_n=\left\{\sum_{i=1}^n (x_i -1)^2 - 4, \, \sum_{i=1}^n (x_i +1)^2 - 4 \right\}$$

The runtimes achieved by the package are the following:

\begin{table}[h]
    \centering
    \begin{tabular}{|c|c|c|}
        \hline
        \textbf{Variables} & \textbf{Runtime (seconds)} & $n$\textbf{-cells} \\ \hline
        1         & 0.0308762         & 5     \\ \hline
        2         & 0.335376          & 29     \\ \hline
        3         & 6.06833           & 467    \\ \hline
        4         & 1883.13           & 7370    \\ \hline
    \end{tabular}
\end{table}

\section{Future work}

Some possible directions for expanding the package include:

\begin{itemize}

    \item \textbf{Improved Output:}
    Enhancing \texttt{liftingPoint} by evaluating the input polynomials at the top-level sample points, providing more information to users and streamlining \texttt{positivePoint}. This would also simplify manual sign-checking of cells and future related commands.

    \item \textbf{Extension to Full CAD:}
    Expanding from Open CAD to a full CAD with cells of every dimension. This would require the implementation of real polynomial division and a way to symbolically store and retrieve real roots, but would enable the following two applications:
    
    \item \textbf{Quantifier Elimination:}
    A full CAD could be used to answer any nonlinear arithmetic quantified questions through quantifier elimination (given sufficient resources).

    \item \textbf{Connectivity Analysis of Algebraic Varieties:}
    By combining a full CAD with an adjacency algorithm, such as the one described in \cite{Strzebonski17}, one could determine the number of connected components in algebraic varieties, offering new insights into their topological properties.
\end{itemize}

These ideas represent potential avenues for further research and development in the domain, aimed at enhancing our understanding and capabilities in computational algebraic geometry.

\newpage

\bibliographystyle{alpha}
\bibliography{bibliography}

\newcommand{\etalchar}[1]{$^{#1}$}
\begin{thebibliography}{HEW{\etalchar{+}}15}

\bibitem[BD07]{BrownDavenport2007}
Christopher~W. Brown and James~H. Davenport.
\newblock {The complexity of quantifier elimination and cylindrical algebraic
  decomposition}.
\newblock In {\em Proceedings of the 2007 International Symposium on Symbolic
  and Algebraic Computation}, ISSAC '07, page 54–60, New York, NY, USA, 2007.
  Association for Computing Machinery.

\bibitem[Bro04]{Brown2004}
C.~W. Brown.
\newblock {ISSAC 2004 Tutorial: Cylindrical Algebraic Decomposition. CADs More
  Formally}.
\newblock
  \url{https://www.usna.edu/Users/cs/wcbrown/research/ISSAC04/slides3F.pdf},
  2004.
\newblock Figure 3, Accessed: 2025-02-28.

\bibitem[Col75]{Collins75}
George~E. Collins.
\newblock {Quantifier elimination for real closed fields by cylindrical
  algebraic decomposition}.
\newblock In H.~Brakhage, editor, {\em Automata Theory and Formal Languages},
  pages 134--183, Berlin, Heidelberg, 1975. Springer Berlin Heidelberg.

\bibitem[CZC20]{ChenEtAl2020}
Changbo Chen, Zhangpeng Zhu, and Haoyu Chi.
\newblock {Variable Ordering Selection for Cylindrical Algebraic Decomposition
  with Artificial Neural Networks}.
\newblock In Anna~Maria Bigatti, Jacques Carette, James~H. Davenport, Michael
  Joswig, and Timo de~Wolff, editors, {\em Mathematical Software -- ICMS 2020},
  pages 281--291, Cham, 2020. Springer International Publishing.

\bibitem[DH88]{DavenportHeintz87}
James~H. Davenport and Joos Heintz.
\newblock Real quantifier elimination is doubly exponential.
\newblock {\em Journal of Symbolic Computation}, 5(1):29--35, 1988.

\bibitem[dRE22]{delRioEngland22}
Tereso del R{\'i}o and Matthew England.
\newblock {New Heuristic to Choose a Cylindrical Algebraic Decomposition
  Variable Ordering Motivated by Complexity Analysis}.
\newblock In Fran{\c{c}}ois Boulier, Matthew England, Timur~M. Sadykov, and
  Evgenii~V. Vorozhtsov, editors, {\em Computer Algebra in Scientific
  Computing}, pages 300--317, Cham, 2022. Springer International Publishing.

\bibitem[EF19]{EnglandFlorescu2019}
Matthew England and Dorian Florescu.
\newblock {Comparing Machine Learning Models to Choose the Variable Ordering
  for Cylindrical Algebraic Decomposition}.
\newblock In Cezary Kaliszyk, Edwin Brady, Andrea Kohlhase, and Claudio
  Sacerdoti~Coen, editors, {\em Intelligent Computer Mathematics}, pages
  93--108, Cham, 2019. Springer International Publishing.

\bibitem[GMSY24]{LopezGarcia2024}
Jordy~Lopez Garcia, Kelly Maluccio, Frank Sottile, and Thomas Yahl.
\newblock {Real solutions to systems of polynomial equations in Macaulay2}.
\newblock {\em Journal of Software for Algebra and Geometry}, 14(1):87--95,
  2024.

\bibitem[GS]{M2}
Daniel~R. Grayson and Michael~E. Stillman.
\newblock Macaulay2, a software system for research in algebraic geometry.
\newblock Available at \url{http://www2.macaulay2.com}.

\bibitem[HEW{\etalchar{+}}15]{HuangEtAl2015}
Zongyan Huang, Matthew England, David Wilson, James~H. Davenport, and
  Lawrence~C. Paulson.
\newblock {A Comparison of Three Heuristics to Choose the Variable Ordering for
  Cylindrical Algebraic Decomposition}.
\newblock {\em ACM Communications in Computer Algebra}, 48(3/4):121–123,
  February 2015.

\bibitem[Jir95]{Jirstrand95}
Mats Jirstrand.
\newblock {Cylindrical Algebraic Decomposition - an Introduction}.
\newblock Technical Report 1807, Linköping University, Department of
  Electrical Engineering, Automatic Control, 1995.

\bibitem[Laz94]{Lazard94}
D.~Lazard.
\newblock {\em {An Improved Projection for Cylindrical Algebraic
  Decomposition}}, volume~26 of {\em Springer Series in Computational
  Mathematics}, pages 467--476.
\newblock Springer New York, New York, NY, 1994.

\bibitem[MPP19]{McCallumEtAl19}
Scott McCallum, Adam Parusiński, and Laurentiu Paunescu.
\newblock {Validity proof of Lazard's method for CAD construction}.
\newblock {\em Journal of Symbolic Computation}, 92:52--69, 2019.

\bibitem[Str00]{Strzebonski00}
Adam Strzeboński.
\newblock {Solving Systems of Strict Polynomial Inequalities}.
\newblock {\em Journal of Symbolic Computation}, 29(3):471--480, 2000.

\bibitem[Str17]{Strzebonski17}
Adam Strzebo\'{n}ski.
\newblock {CAD Adjacency Computation Using Validated Numerics}.
\newblock In {\em Proceedings of the 2017 ACM on International Symposium on
  Symbolic and Algebraic Computation}, ISSAC '17, page 413–420, New York, NY,
  USA, 2017. Association for Computing Machinery.

\bibitem[Tar51]{Tarski51}
Alfred Tarski.
\newblock {\em {A Decision Method for Elementary Algebra and Geometry: Prepared
  for Publication with the Assistance of J.C.C. McKinsey}}.
\newblock RAND Corporation, Santa Monica, CA, 1951.

\end{thebibliography}

\newpage

\end{document}